\documentclass[showpacs,superscriptaddress,amsmath,aps]{revtex4}
\usepackage{graphicx,color}
\usepackage{bm}
\usepackage[hypertex]{hyperref}

\newcommand{\be}{\begin{equation}}
\newcommand{\ee}{\end{equation}}
\newcommand{\bea}{\begin{eqnarray}}
\newcommand{\eea}{\end{eqnarray}}
\newcommand{\bsube}{\begin{subequations}}
\newcommand{\esube}{\end{subequations}}

\newcommand{\Eq}[1]{Eq.\,(\ref{#1})}

\newcommand{\la}{\langle}
\newcommand{\ra}{\rangle}

\newcommand{\nl}{\nonumber \\}

\newcommand{\beq}{\begin{equation}}
\newcommand{\eeq}{\end{equation}}
\newcommand{\beqn}{\begin{eqnarray}}
\newcommand{\eeqn}{\end{eqnarray}}
\newcommand{\bsub}{\begin{subequations}}
\newcommand{\esub}{\end{subequations}}

\begin{document}


\title{ Revisit the spin-FET: Multiple reflections,
inelastic scattering, and lateral size effects}

\author{Luting Xu}
\affiliation{Department of Physics, Beijing Normal University,
Beijing 100875, China}
\author{Xin-Qi Li}
\email{lixinqi@bnu.edu.cn}
\affiliation{Department of Physics, Beijing Normal University,
Beijing 100875, China}
\author{Qing-feng Sun}
\affiliation{International Center for Quantum Materials,
School of Physics, Peking University, Beijing 100871, China}

\date{\today}

\begin{abstract}
We revisit the spin-injected field effect transistor (spin-FET)
by simulating a lattice model
based on recursive lattice Green's function approach.
In the one-dimensional case and coherent regime,
the simulated results reveal noticeable differences
from the celebrated Datta-Das model, which motivate thus
an improved treatment and lead to analytic and generalized result.
The simulation also allows us to address inelastic scattering
(using B\"uttiker's fictitious reservoir approach)
and lateral confinement effects on the control of spins
which are important issues in the spin-FET device.
\end{abstract}


\maketitle

\section{Introduction}

The spin-valve device \cite{Ju75,Slon89,Mood95}
and spin-injected field effect
transistor (spin-FET) \cite{DD90} lie at the heart of spintronics.
The basic principle of this type of devices is modulating the resistance
by controlling the spins of the carriers \cite{Wolf01,Zu04},
in particular employing two ferromagnetic (FM) leads
as polarization generator and detector.
In practice, there existed two major challenges:
(i) spin-polarized injection into a semiconducting channel, and (ii)
gate control of the Rashba spin-orbit coupling (SOC) in the channel.
The former difficulty has been largely overcome through
efforts of many groups \cite{Rash00,Fert01,Smi01,Bau05}.
For the latter issue, investigations included
the gate-voltage-controlling of the spin precession in
both the quantum wells \cite{Nit97} and quantum wires \cite{Eng97},
and some detailed studies such as the
multichannel mixing effects (lateral size effects)
\cite{Pa02,Go04,Ni05,Je06,Liu06,Ge10,Pala04,Ag10}.
Integrating the ingredients of the two types mentioned above
into a single device using AsIn heterostructure with a top gate
was realized in a recent experiment \cite{KK09}.
This progress, remarkably, has renewed the interest
in the spin-FET device \cite{Ge10,Ag10,Ge10a,Dat11},
which was proposed for some time
longer than two decades by Datta and Das \cite{DD90}.

In this work we revisit this novel spintronic device,
based on the powerful recursive lattice
Green's function (GF) simulation on
a quantum-wire model (semiconductor nanowire implementation).
To reach realistic scales, from the InSb material parameters
(which have large Land\'e $g$ factor
and strong spin-orbit coupling \cite{Kou12}),
we design our simulation size (in longitudinal direction)
for the quantum wire with 500 lattice sites (about $300$ nm length).
We may summarize the present study to step the following advances:
(i)
In the ideal one-dimensional (1-D) case and coherent regime,
the simulated results of the energy-resolved
transmission spectrum and the SOC-modulation
of the transmission peak reveal interesting differences
from the well-known Datta-Das model \cite{DD90}.
Accordingly, we develop a Fabry-Perot cavity model
to obtain an analytic result which generalizes
Ref.\ \cite{DD90}, and as well the more recent work \cite{Dat11}.
(ii)
The employed recursive GF technique allows for an efficient
simulation for the spatial-motion decoherence effect which,
quite {\it indirectly}, degrades the control of the spin precession.
Of particular interest is that this treatment
does not involve any explicit spin-flip mechanisms \cite{Dya7172},
but only incorporates the B\"uttiker phase-breaking model
\cite{But8688,Pas90,Li01}
to introduce spatial decoherence effect.
The simulated result agrees with the temperature dependence
observed in experiment \cite{KK09}, and substantiates
the mesoscopic (coherence) requirement remarked
in the Datta-Das proposal \cite{DD90} or
the non-diffusive (ballistic) criterion \cite{Eom11}.
(iii)
We simulate the effect of lateral confinement by setting
20 and 40 lattice sites for the width of the quantum wire.
The results are in consistence with some previous studies
based on continuous wave-guide models
\cite{Pa02,Go04,Ni05,Je06,Liu06,Ge10,Pala04,Ag10},
implying that the lateral size, if exceeding certain range
(drastically violating the 1-D condition),
will influence the functionality of the spin-FET device.

\section{Model and Methods }

The device contains a central region (quantum wire)
and two FM leads, described by total Hamiltonian
$H=H_{\rm nw}+\sum_{\beta=L,R}H_{\beta}+H_T$, with
\begin{subequations}\label{Hsys}
\begin{align}
& H_{\rm w}=\sum_{i=1}^M \epsilon_{\tiny 0} c^{\dag}_ic_i
-\sum_{i=1}^{M-1} [(t_0 c^{\dag}_{i+1}c_{i} \nonumber\\
& ~~~~~~~~~~~~
-i\alpha c_{i+1}^{\dag}\sigma_y c_{i})+ {\rm H.c.} ] \;,  \\
\nonumber\\
& H_{\beta}=\sum_{i=1}^{\infty} ~ b_{\beta,i}^{\dag}
(\epsilon_{\beta} +\bm{\sigma\cdot h}_{\beta}) b_{\beta,i}  \nonumber\\
& ~~~~~~~
- \sum_{i=1}^{\infty}(t_{\beta} b^{\dag}_{\beta, i+1}
   b_{\beta,i} + {\rm H.c.} ) \;,\\
\nonumber\\
& H_T = -t_c[(c_1^{\dag} b_{L,1}+c^{\dag}_M b_{R,1} )+{\rm H.c.} ] \;.
\end{align}
\end{subequations}
%
For the sake of simplicity, here we specify these Hamiltonians using
1-D tight-binding lattice model (with $M$ lattice sites for the quantum wire),
and will present extra explanations
when extended to higher dimensions (in Sec.\ III C).
The electronic creation and annihilation operators
are abbreviated by a vector form, e.g.,
$c_i^{\dag}=(c_{i\uparrow}^{\dag}, c_{i\downarrow}^{\dag})$
and $b_{\beta, i}^{\dag}=(b_{\beta, i\uparrow}^{\dag},
b_{\beta, i\downarrow}^{\dag})$,
where $i$ labels the lattice site
and $(\uparrow,\downarrow)$ the spin orientations.
The Pauli matrices are introduced as
$\bm{\sigma}=(\sigma_x,\sigma_y,\sigma_z)$.
In the quantum wire Hamiltonian ($H_{\rm w}$),
$\epsilon_{\tiny 0}$ and $t_0$
are the tight-binding site energy and hopping amplitude;
$\alpha$ is the SOC strength.
Notice that, when converting to a continuous model,
the corresponding SOC strength should be
$\widetilde{\alpha}=2a\alpha$, where $a$ is the lattice constant.
For the FM leads ($H_{\beta}$),
$\epsilon_{\beta}$, $t_{\beta}$,
and $\bm{h}_{\beta}$ are, respectively,
the tight-binding parameters and the FM exchange field.
For the wire-lead coupling ($H_T$), we assume
a common coupling amplitude $t_c$ at both sides.

In more detail, the FM exchange field $\bm{h}_{\beta}$
takes the direction of magnetization.
For spintronic device such as the spin-valve,
the left and right leads should have different
magnetization directions,
and the device function is realized by tuning one of them.
However, for the spin-FET,
whose function is tuned by manipulating
the spin precession in the central region,
we can assume the FM leads magnetized in parallel, e.g.,
with $\bm{h}_{\beta}=h_0 (0,0,1)$ for a $z$-axis magnetization.

\subsection{Inelastic Scattering Model}

In this work we will address the important issue of decoherence
(inelastic scattering) effect in the spin-FET.
Rather than the electron-phonon interactions, which are difficult
to treat in large-scale simulation of quantum transports,
we would like to employ the simpler but somehow equivalent
phenomenological phase-breaking approach
proposed by B\"uttiker \cite{But8688}.

The basic idea of this approach
is to attach the system (quantum wire) to some
additional {\it virtual} electronic reservoirs.
The transport electron is assumed to partially enter the virtual reservoir,
suffer an inelastic scattering in it (then lose the phase information),
and return back into the system
(to guarantee the conservation of electron numbers).
As a consequence, the two partial waves of electron, say,
the component that once entered the reservoir and the one having not,
do not interfere with each other.
Technically, we model the virtual reservoir
(coupled to the $J_{\rm th}$ site of the quantum wire)
by a tight-binding chain with Hamiltonian \cite{Pas90,Li01}
\bea
\widetilde{H}_{J}= \sum^{\infty}_{i=1} \epsilon_{\tiny 0}
b^{\dag}_{J,i} b_{J,i}
- \sum^{\infty}_{i=1}
(t_J b^{\dag}_{J,i+1} b_{J,i}+{\rm H.c.} )\, ,
\eea
and this chain is coupled to the quantum wire through
a coupling Hamiltonian
\bea
\widetilde{H}_{T,J}=-(\eta c^{\dag}_J b_{J,1} + {\rm H.c.}) \,,
\eea
with $\eta$ the coupling strength.
In this work, for the quantum wire with $M=500$ (length of $\sim 300$ nm),
we will assume to attach 10 side-reservoirs (so $J=50, 100, \cdots, 500$),
which represent a mean-distance of $50 a\simeq 30$ nm
between the nearest-neighbor inelastic scatterers.
This mean-distance and the coupling strength ($\eta$),
jointly, characterize the decoherence strength \cite{But8688}.

\subsection{Lattice Green's Function Method}

We will base our simulation on the powerful
lattice Green's function (GF) method,
in particular combined with a {\it recursive} algorithm \cite{Datta95}.
For large-scale simulation, this technique can avoid
using all the lattice sites as state basis,
needing only a piece of the lateral lattice sites
for a matrix representation.
The longitudinal lattice sites are
treated by a recursive algorithm.
The great advantage of this treatment is saving
the dimension of the representation matrix.

Based on the recursive algorithm, one can calculate the
retarded and advanced Green's functions and obtain the transmission
coefficients between any pair of leads (reservoirs)
as follows \cite{Datta95,QF01,QF09}:
\begin{equation}\label{Tmn}
T_{\mu\nu}(\epsilon)={\rm Tr}[\Gamma_{\mu}(\epsilon)G^r(\epsilon)
\Gamma_{\nu}(\epsilon)G^a(\epsilon)] .
\end{equation}
Here we use $\mu$ ($\nu$) to denote all the reservoirs,
including the left and right leads
together with the virtual inelastic scattering reservoirs.
Formally, $\Gamma_{\mu}(\epsilon)
=i[\Sigma^r_{\mu}(\epsilon)-\Sigma^a_{\mu}(\epsilon)]$,
and $G^r(\epsilon)=[G^a(\epsilon)]^{\dag}
=1/[\epsilon-H_{\rm WR}-\sum_{\mu=L,R,\{J\}}(\Sigma_{\mu}^r)]$.
$\Sigma_{\mu}^{r(a)}$ is the retarded (advanced)
self-energy owing to coupling with the $\mu_{\rm th}$ lead (reservoir).
In practice, $\Sigma_{\mu}^{r(a)}$ can be easily obtained
by surface Green's function technique,
and the full-system Green's function $G^{r(a)}$
can be efficiently computed using the recursive algorithm.

Knowing $T_{\mu\nu}$, the entire {\it effective transmission coefficient}
from the left to the right lead can be straightforwardly
obtained through \cite{Pas90,Li01}
\begin{equation}\label{Teff}
{\cal T}_{\rm eff}(\epsilon)= T_{LR} + \sum^{N}_{\mu,\nu=1}
K^{(L)}_{\mu} W^{-1}_{\mu\nu} K^{(R)}_{\nu} .
\end{equation}
Here, $K^{(L)}_{\mu}=T_{L\mu}$ and $K^{(R)}_{\nu}=T_{\nu R}$.
$W^{-1}$ is the inverse of the matrix $W$ with elements
$W_{\mu\nu}=(1-R_{\nu\nu})\delta_{\mu\nu}
-T_{\mu\nu}(1-\delta_{\mu\nu})$,
where $R_{\nu\nu}=1-\sum_{\mu(\neq\nu)}T_{\nu\mu}$.
Inserting the transmission coefficient ${\cal T}_{\rm eff}(\epsilon)$
into the Landauer-type formula,
one can easily compute the transport current.
In this work, however, we will simply use
${\cal T}_{\rm eff}(\epsilon)$
(corresponding to differential conductance)
to characterize the modulation effects in the spin-FET.

\section{Results and Discussions}

In our simulation, for the central quantum wire,
we refer to the SOC strength of the InSb material,
$\widetilde{\alpha} = 0.2 ~{\rm eV}{\cdot}{\rm \AA}$.
This implies a SOC length $l_{so}\simeq 200~{\rm nm}$.
Assuming a lattice constant $a\simeq 6~{\rm \AA}$,
we then decide to simulate the 1-D quantum wire with $M=500$ lattice sites
(length of $\sim 300$ nm),
in order to be longer than $l_{so}$ for the purpose of spin-FET.
For the tight-binding hopping energy, we assume $t_0= 1.0$ eV.
For the FM leads and the fictitious (inelastic scattering) reservoirs,
we assume common hopping parameter in their tight-binding models,
i.e., $t_{\beta}=t_J=0.8$ eV.
Finally we assume splitting exchange energy
$h_{0} = 0.4$ eV for the FM leads,
and $t_c = 0.4 t_0$ for their coupling to the quantum wire.

Let us consider first a {\it coherent} transport through the quantum wire
(corresponding to the case of low temperatures \cite{KK09}).
In Fig.\ 1(a), we display the representative results of the
transmission spectrum under the SOC ($\alpha$) modulation.
Owing to finite length of the quantum wire, the transmission
spectrum reveals the usual peak-{\it versus}-valley structure.
The SOC-induced energy level splitting also
results in additional fine-structures
(see, for instance, the red curve).
In Fig.\ 1(a) we observe clear SOC-modulation
effect on the entire transmission spectrum.
For convenience but without loss of physics,
in this work we would like to employ the height
of the transmission peak to characterize
the modulation effect.
The extracted results are shown in Fig.\ 1(b).

We find that the SOC-modulation period is well described
by $\alpha^*=\pi t_0/M$, where $M=L/a$.
This is in perfect agreement with the result from a simple
plane-wave-based interference analysis.
Following Ref.\ \cite{DD90},
the phase difference caused by the SOC over distance $L$
between the spin-up and spin-down components is given by
$\theta=(2m^*/\hbar^2)\widetilde{\alpha} L$.
In order to convert to the lattice model,
making replacement $\hbar^2/2m^* \rightarrow t_0 a^2$
(and noting that $\widetilde{\alpha}=2a\alpha$),
the above $\alpha^*$ is then given by the condition $\theta=2\pi$.

However, the SOC-modulation lineshape does not coincide in general
with the prediction of the Datta-Das model \cite{DD90}.
In Ref.\ \cite{DD90}, it was remarked that
the SOC-modulation effect is free from energies.
However, as we will prove shortly, this is not true.
Also, we find different transition behaviors
around the (modulation) peaks and valleys:
the variation around the peak can be much slower
(forming almost a ``plateau") than the change around the valley.
Below we present a semi-quantitative analysis
based on essentially the same Datta-Das model
but accounting for {\it multiple reflections}
in the SOC region, which desirably
generalizes the central result in Ref.\ \cite{DD90}.
We notice also that this type of multiple reflections
were not taken into account when fitting and
analyzing the experimental result \cite{KK09,Eom11,Dat11}.

\begin{figure}
  \centering
  \includegraphics[scale=0.7]{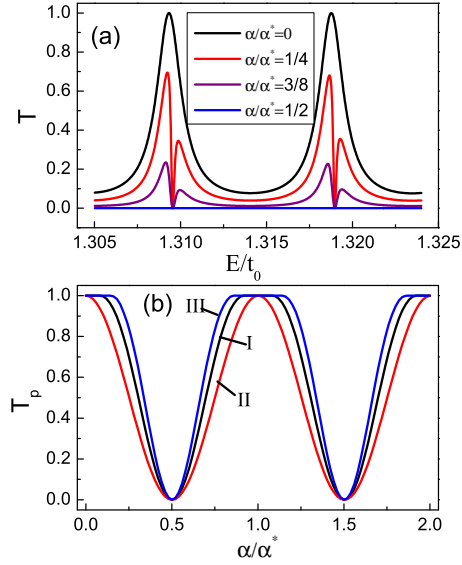}\\
  \caption{(color online)
(a)
  SOC($\alpha$)-modulation effect on the transmission spectrum,
  with modulation period $\alpha^*=\pi t_0/M$
  (see the main text for more detail).
  In particular, at $\alpha=0.5\alpha^*$
  the entire transmission spectrum is suppressed,
  indicating an {\it off}-state of the spin-FET.
(b)
  SOC($\alpha$)-modulation to the heights of the transmission peaks.
  Illustrative results are shown for three energy intervals:
  curve I for $E/t_0\in(1.305,1.315)$,
  II for $E/t_0\in(1.45,1.46)$,
  and III for $E/t_0\in(1.70,1.71)$.  }
\end{figure}

\begin{figure}
  \centering
  \includegraphics[scale=0.7]{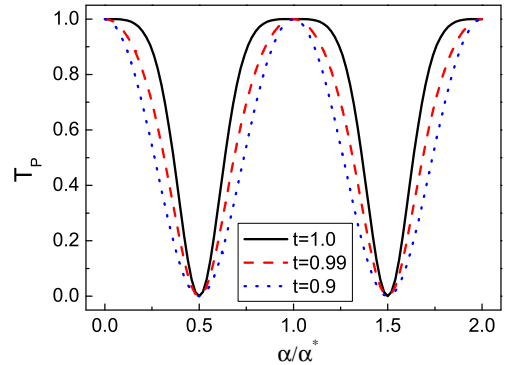}
  \caption{(color online)
  SOC($\alpha$)-modulation effect from a Fabry-Perot-type
  resonator model consideration, in which
  the multiple reflections are essentially accommodated.
  The (resonant) transmission peak ($T_p$) is obtained from
  \Eq{T-model} and a ``plateau" behavior is recovered
  when the single-side transmission is nearly transparent
  (the transmission coefficient $t\simeq 1$).
  However, along the decrease of $t$, the result
  will be soon close to the one without accounting for
  the multiple reflections \cite{DD90}
  (see main text for more detailed explanation).   }
\end{figure}

\subsection{Semi-quantitative Analysis}

Let us consider a 1-D continuous model
for the quantum wire embedded in between two FM leads.
This is similar to an optical two-sided Fabry-Perot cavity system,
with the electron transmission as an analog of optical wave.
Of particular interest in the electronic setup of spin-FET is the
SOC-modulation in the ``cavity",
which is described by the continuous version of the Rashba model as
$H_{\rm so}=\widetilde{\alpha} (\sigma_x k_y-\sigma_yk_x)
\equiv -\widetilde{\alpha} k\sigma_y$,
owing to the 1-D motion with $k_y=0$.
As in the lattice model, we assume the FM leads
polarized in $z$-direction.
Following the analysis of Datta and Das \cite{DD90},
the state of the injected electron is decomposed
in superposition of the $\sigma_y$-eigenstates:
$|\psi_1\ra=a |\uparrow\ra_y + b |\downarrow\ra_y$
(actually $a=b=1/\sqrt{2}$ in this special case).
Then, after a single passage through the (SOC) 1-D wire
(forward propagation over distance $L$), the state evolves to
$|\psi_2\ra=ae^{ik_+L}|\uparrow\ra_y + be^{ik_-L}|\downarrow\ra_y$.
Here $k_{\pm}$ are given by the solution from
$E=\hbar^2k^2/2m^* \mp \widetilde{\alpha} k $,
for a given energy $E$ \cite{note-1}.

From a different perspective, in the $\sigma_z$ representation,
this evolution manifests an effect of spin precession with angle
$\theta=(k_+-k_-)L=(2 m^*/\hbar^2) \widetilde{\alpha} L$.
Taking into account the role of the FM leads,
a transmission coefficient was proposed
in Ref.\ \cite{DD90} as $T \propto \cos^2(\theta/2)$.
In practice \cite{KK09}, this result has been applied
to analyze experiment as follows: the measurement voltage,
which is proportional to the transmission coefficient,
is fitted with $V=A\cos(2 m^*\widetilde{\alpha} L/\hbar^2+\varphi)$,
where $A$ and $\varphi$ are two fitting parameters.
In a more recent work \cite{Dat11}, deeper analysis was carried out
for these two parameters (amplitude and phase),
and some aspects of the experiment were explained
while some others remained unclear.

A drawback in the above analysis is the neglect
of the (infinitely) multiple reflections, which actually
exist in any two-leads connected electronic devices.
In terms of wave-mechanics treatment, this is exactly
the same as the optical two-sided Fabry-Perot cavity.
In addition to the forward propagation (spin precession caused by the SOC),
one can similarly account for the spin precession
in the backward propagation
(after reflection at the junction connected with the lead).
Particularly, as accounting for the multiple reflections,
one should adopt a {\it full} reflection for the anti-parallel
spin component (with respect to the FM polarization),
and the usual transmission and reflection
for the component of parallel orientation.
As a result, after each reflection at the junction
connected with the FM lead, the reflected electron
would suffer an amount of spin rotation.
After some algebra (summarized in Appendix A),
the final result reads
\bea\label{T-model}
T= \frac{4t^4\cos^2(\theta/2) \sin^2(KL)}
{D^2 + 4t^4\cos^2(\theta/2) \sin^2(KL)}  \,.
\eea
In this result we have denoted
$(r-1)^2\sin^2(\theta/2)+4r\sin^2(KL)\equiv D$
and $(k_++k_-)L/2\equiv KL$.
For the contact of the quantum wire with the FM leads,
we assumed identical transmission ($t$)
and reflection ($r$) amplitudes at the two sides.
We may remark that, after accounting for the (infinite)
multiple reflections, \Eq{T-model}
generalizes the result of Ref.\ \cite{DD90}, elegantly.

Based on \Eq{T-model} we show in Fig.\ 2 the SOC-modulation
effect on the (resonant) transmission peak.
Interestingly, we find similar lineshape as in Fig.\ 1(b).
In particular, different transition behaviors
are found around the peak (maximum) and dip (minimum),
for the case $t\simeq 1$.
From \Eq{T-model} and setting $t=1$, we obtain
$T_p=4\cos^2(\theta/2)/[1+\cos^2(\theta/2)]^2$.
This result, in a simple way, allows us to explain
the ``plateau" behavior of $t\simeq 1$ in Fig.\ 2.

We find in Fig.\ 2 that, with the decrease of $t$,
the transmission peak modulation given by \Eq{T-model}
approximately coincides with the {\it energy-independent}
modulation predicted in Ref.\ \cite{DD90}.
This feature should deserve particular attention,
since it may mask the effect of multiple reflections.
From \Eq{T-model}, under the condition
$[(r-1)^2/(4r)]\sin^2(\theta/2)\leq 1$,
an extremal analysis gives the height
of the transmission peak as
\bea\label{Tp}
T_p =\left[ 1-\frac{(r-1)^2}{(r+1)^2} \sin^2\left(\frac{\theta}{2}\right)
     \right]^{-1} \cos^2\left(\frac{\theta}{2}\right) \,.
\eea
We see that, with the decrease of $t$ (from unity),
$T_p$ will soon be close to $\cos^2(\theta/2)$.
For instance, for the still relatively large $t=0.9$,
one can check $(r-1)^2/(r+1)^2\simeq 0.15$.
This will make the effect of the second term
in the square brackets in \Eq{Tp} negligible
(when $t$ is smaller than certain values),
as observed in Fig.\ 2.

Shown in Fig.\ 2 is only the SOC-modulation effect on the transmission peak.
For the entire transmission spectrum, given by \Eq{T-model},
it is clear that the modulation effect depends on energies,
through the $K$-dependence.
This will more dramatically affect the finite-bias current
through the spin-FET, compared with
the {\it energy-independent} modulation effect \cite{DD90}.
In Ref.\ \cite{DD90}, it was highlighted that
the {\it energy-independent} modulation ``property",
observed from the differential phase shift
$\theta=2m^*\widetilde{\alpha} L/\hbar^2$,
implies an important advantage
for quantum-interference device applications.
That is, it can avoid washing out the interference effects and achieve
large percentage modulation of the current, even in multimoded
devices operated at elevated temperatures and large applied bias \cite{DD90}.
It seems of interest to perform further examination
on these statements based on \Eq{T-model}.

Using \Eq{T-model}, one may qualitatively understand
the modulation behavior in Fig.\ 1(b).
Compared the lattice system described by \Eq{Hsys}
with the Fabry-Perot cavity model,
an obvious difference is that the former does not have
a constant single-side transmission ($t$) and reflection ($r$),
where the effective $t$ and $r$ should depend on the energy ($E$)
and the SOC $\alpha$.
Therefore, from \Eq{T-model}, the transmission peak $T_p$
may have different height in different energy region
and may depend on $\alpha$ through the effective $t$ and $r$.
In addition to the multiple reflections,
this should be the reason that lead to the non-overlapped
modulation lineshapes in different energy areas
and the ``plateau" behavior around the modulation peak,
as shown in Fig.\ 1(b).

\subsection{Decoherence Effect}

In Fig.\ 3 we show decoherence effect on the SOC-modulation,
using the B\"uttiker phase-breaking model
as briefly outlined in Sec. II A.
This phenomenological approach is very efficient
compared to any other microscopic model based treatments.
However, from \Eq{Teff}, we see that we need to calculate
all the $T_{\mu\nu}$, based on \Eq{Tmn}.
For each $T_{\mu\nu}$, we need to
recursively calculate the (full system) Green's function
from the $\mu_{\rm th}$ reservoir to the $\nu_{\rm th}$ one.
It will be very computationally expensive.
In practice, however, one can design smart algorithm
to avoid this type of repeated recursive computations.

Qualitatively speaking, the inelastic scatterers would
cause a large number of forward and backward propagation pathways.
Simple analysis in terms of time-reversal symmetry tells us that
the forward and backward propagation over equal distance
would cancel the spin precession.
As a result, for any transmitted electron
(from the left to the right leads),
the {\it net} distance of spin precession is the length of the quantum wire.
This explains the common SOC-modulation period ($\alpha^*$)
in Fig.\ 3 when altering the inelastic scattering strength ($\eta$).

However, the SOC-modulation amplitude will be suppressed
by enhancing the inelastic scattering strength.
The fictitious reservoir model is very convenient to account for
phase breaking (decoherence) of spatial motion,
through destroying quantum interference between partial waves.
Nevertheless, to the SOC caused spin precession,
the role of this model is not so straightforward.
We may remark that in our treatment we did not introduce
{\it explicit} spin-relaxation mechanism \cite{Dya7172},
whose effect is relatively more direct \cite{Dat11,Eom11}.
In Ref.\ \cite{DD90}, Datta and Das pointed out that,
in order to perform the spin-FET,
one of the essential requirements is the central conducting
channel within a mesoscopic phase-coherent regime.
Our result in Fig.\ 3 substantiates this requirement,
and as well the general remark that the Rashba-spin-control
does not work in diffusive transport regime \cite{Eom11}.
The present result is also in agreement with the experiment \cite{KK09},
where the SOC modulation effect was found to be washed out
with stronger inelastic scattering (more phonon excitations)
by increasing the temperatures.

\begin{figure}
  \centering
  \includegraphics[scale=0.7]{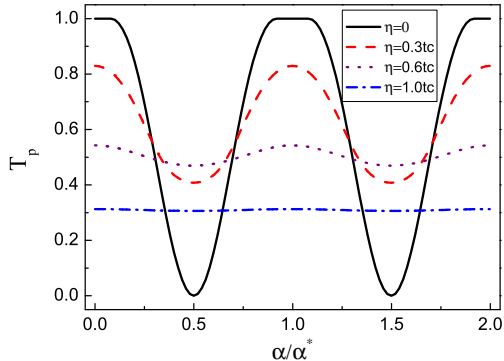}\\
  \caption{(color online)
  Decoherence effect on the SOC($\alpha$)-modulation displayed
  in Fig.\ 1(b), for the transmission peak with $E/t_0\in(1.305,1.315)$.
  The coupling coefficient ($\eta$) to the fictitious side-reservoirs
  characterizes well the decoherence strength in the B\"uttiker
  phase-breaking approach. }
\end{figure}

\subsection{Lateral-Size Effect}

In the original proposal
the intersubband coupling effect owing to lateral size
was excluded for a narrow (quasi-1D) quantum wire \cite{DD90}.
Below, employing the recursive lattice GF approach,
we simulate the lateral-size effect
by considering a quasi-two-dimensional (2D)
quantum ribbon with $M\times N$ lattice sites.
Accordingly, we need to generalize each lattice site of the 1-D wire
to a lateral column with $N$ sites along the $y$-direction.
While the 2D generalization of the tight-binding model is straightforward,
we only specify the SOC Hamiltonian in 2D case as
\bea
H_{SO}= \sum_{i}
\left[ i\alpha \left( a^{\dagger}_i \sigma_y a_{i+\delta x}
   - a^{\dagger}_i \sigma_x a_{i+\delta y}\right)
   +{\rm H.c.} \right]  \,.
\eea
Here, the summation is over the $M\times N$ lattice sites,
and $(\delta_x,\delta_y)=(1,1)$
denote displacements over a unit lattice cell
along the longitudinal ($x$) and lateral ($y$) directions.

\begin{figure}
  \centering
  \includegraphics[scale=0.7]{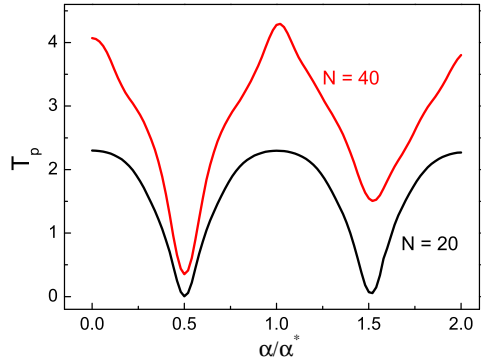}
  \caption{(color online)
   Effect of the confined lateral motion on the SOC($\alpha$)-modulation.
   Within the lattice model,
   in addition to the longitudinal number of sites $M=500$,
   we set $N=20$ and $40$ to reveal the increasingly non-negligible
   lateral size effect \cite{note-2}.   }
\end{figure}

In Fig.\ 4 we show the effects of the lateral size (with $N=20$ and $40$).
First, owing to the energy sub-bands (and their mixing)
caused by the lateral confinement, the transmission peak
can exceed unity (in the 1-D case the maximal $T_p$ is unity).
One may notice that, unlike prediction from the standard
Landauer-B\"uttiker formula, the height of the transmission peak,
which is proportional to the differential conductance
at the corresponding energy (bias),
does not equal the lateral-channel numbers involved.
In 1-D case, the transmission peak is originated from a constructive
interference given by the standing-wave condition
of the longitudinal wave-vector.
This resonant condition (together with symmetric coupling
to the leads) will result in a transmission coefficient of unity.
However, for a given energy ($E$) in the quantum-ribbon system,
different lateral channels are associated with
different longitudinal wave-vectors.
Then, the resonant condition for each transverse channel
cannot be satisfied simultaneously.

The second effect originated from the lateral motion
is the SOC-induced {\it additional} spin precession.
In general, this will affect the SOC-modulation
quality of the spin-FET.
For small $N$ (with respect to the SOC length with $\sim 300$
sites in the present study), this effect is not prominent
(see, for instance, the result of $N=20$ in Fig.\ 4).
However, with the increase of the lateral size,
the transmission cannot be switched off
(particularly at higher $\alpha$),
as illustrated by the result of $N=40$.
Moreover (not shown in Fig.\ 4),
with even larger lateral-size and SOC-$\alpha$,
or in some energy domain, the transmission modulation
will become strongly irregular.
We then conclude that, while the longitudinal modulation
period ($\alpha^*$) keeps unchanged,
the quality of the spin-FET performance will be degraded
with the increase of the lateral size.
Only for narrow quantum wire (small $N$) and relatively weak $\alpha$,
one can define desirable working region for the spin-FET.
This remark supports the conclusion in Ref.\ \cite{Ge10},
and some previous studies \cite{Pa02,Go04,Ni05,Je06,Liu06}.
It seems that an exception is the 2D system
with semi-infinite (considerably wide) width,
where the SOC modulation effect,
despite of the degraded quality, can be restored
\cite{Pala04,Ag10,Dat11,Eom11}.


\section{Summary}

We have revisited the transport
rooted in the spin-FET device,
with the help of the powerful recursive
lattice Green's function approach.
Our result of the energy-resolved transmission spectrum reveals
noticeable differences from the Datta-Das model \cite{DD90},
which motivated us to develop a Fabry-Perot-cavity type
treatment to generalize the central result.
We also simulated the decoherence and lateral-size effects.
The former substantiates the mesoscopic
(coherence) requirement \cite{DD90}
or the non-diffusive (ballistic) criterion \cite{Eom11},
and is in reasonable agreement with the
observation in the recent experiment \cite{KK09}.
The latter implies additional restrictions to
the Rashba-spin-control and thus the quality of the device.

\vspace{0.8cm}
{\it Acknowledgments}---
This work was supported by the Major State Basic Research
Project of China (Nos.\ 2011CB808502 \& 2012CB932704)
and the NNSF of China (No.\ 91321106).

\appendix

\section{Multiple Reflection Analysis}

Following the Fabry-Perot cavity model explained
in Sec.\ III A, let us consider the transmission
and reflection of an electron at the right FM lead,
which entered from the left FM lead with a wave function
(after passing through the left contact junction):
\begin{equation}
|\psi_1\rangle=a|\uparrow\rangle_y+b|\downarrow\rangle_y .
\end{equation}
At the right side of the 1-D wire (cavity),
after single passage (over $L$) under the SOC influence,
the electron state evolves to
\begin{eqnarray}
|\psi_2\rangle &=& a e^{ik_+L}|\uparrow\rangle_y
    +b e^{ik_-L}|\downarrow\rangle_y\nonumber\\
&=&\left(
\begin{array}{cc}
e^{ik_+ L}&0\\
0&e^{ik_- L}
\end{array}
\right)|\psi_1\rangle
\end{eqnarray}
Here and in the following, using the transfer matrix representation,
the states should be understood as column vectors in the basis
$\{ |\uparrow\rangle_y, |\downarrow\rangle_y \}$,
e.g., $|\psi_1\rangle=(a,b)^T$.
For a given energy $E$, $k_{\pm}$ are solved from
$E=\hbar^2k^2/2m^* \mp \widetilde{\alpha} k$,
corresponding to the momentums of the spin-up
and spin-down electrons.

Since the FM leads are polarized in the $z$-direction,
at the right side,
only the electron with spin state $|\uparrow\rangle_z$
can enter the right lead (with transmission amplitude $t$
and reflection amplitude $r$).
For electron with $|\downarrow\rangle_z$,
it will be {\it fully} reflected.
Accordingly, based on $|\psi_2\rangle$,
the transmitted wave into the right lead is given by
\begin{eqnarray}\label{psi1}
|\psi\rangle^{(1)}_R
= t \hat{P}_{z\uparrow} |\psi_2\rangle
\equiv U_{R} |\psi_1\rangle .
\end{eqnarray}
In this context we introduce the projection operators
\bea
\hat{P}_{z\uparrow(\downarrow)}
=|\uparrow(\downarrow)\ra_z \la \uparrow(\downarrow)| \,.
\eea
In \Eq{psi1} we also defined a transfer matrix which reads
\begin{eqnarray}
U_{R}=\frac{t}{2}\left(
\begin{array}{cc}
e^{ik_+L}&e^{ik_-L}\\
e^{ik_+L}&e^{ik_-L}
\end{array}
\right) \,.
\end{eqnarray}
At the same time, the reflected wave from the right junction
is given by
\begin{eqnarray}
|\widetilde{\psi}\rangle^{(1)}_R
= \hat{P}_{z\downarrow}|\psi_2\rangle
   + r \hat{P}_{z\uparrow} |\psi_2\ra
\equiv \widetilde{U}_{R} |\psi_1\rangle \,,
\end{eqnarray}
where
\begin{eqnarray}
\widetilde{U}_{R}=\frac{1}{2}\left[
\begin{array}{cc}
(r+1)e^{ik_+L}&(r-1)e^{ik_-L}\\
(r-1)e^{ik_+L}&(r+1)e^{ik_-L}
\end{array}
\right] \,.
\end{eqnarray}
Similar analysis gives the transfer matrix acting on the wave
inversely propagated from the right side to the left one
and reflected at the left junction:
\begin{eqnarray}
\widetilde{U}_{L}=\frac{1}{2}\left[
\begin{array}{cc}
(r+1)e^{ik_-L}&(r-1)e^{ik_+L}\\
(r-1)e^{ik_-L}&(r+1)e^{ik_+L}
\end{array}
\right]   \,.
\end{eqnarray}
Therefor, the total wave arriving to the right FM lead
is a sum of all the partial waves, given by
\bea
|\Psi\rangle_R
&=& |\psi\rangle_R^{(1)}+|\psi\rangle_R^{(2)}
    +|\psi\rangle_R^{(3)}+\cdots  \nl
&=& (U_R+U_R\widetilde{U}_L\widetilde{U}_R+\cdots)|\psi_1\rangle \nl
&=& U_R(1-\widetilde{U}_L\widetilde{U}_R)^{-1} |\psi_1\rangle   \,.
\eea
Noting that $|\psi_1\rangle=t|\uparrow\rangle_z$,
we finally obtain the total transmission probability as
\begin{eqnarray}
T=|_z\langle \uparrow |\Psi\rangle_R|^2
= \frac{4t^4\cos^2(\theta/2) \sin^2KL}
{D^2+4t^4\cos^2(\theta/2) \sin^2KL }   \,.
\end{eqnarray}
Here we defined
$KL=(k_++k_-)L/2$ and $\theta=(k_+-k_-)L$,
and introduced
$D=(r-1)^2\sin^2(\theta/2)+4r\sin^2KL$.


\begin{references}
\bibitem{note-1}
For an 1-D lattice model, $k_{\pm}$ can be solved from
$E=[\epsilon_0-2t_0\cos(ka)]\mp 2\alpha\sin(ka)$,
giving $k_+-k_-=(a/2)\arctan(\alpha/t_0)$.
This result indicates, as the same as the 1-D continuous model,
that the spin precession angle $\theta=(k_+-k_-)L$
is independent of the energy.
\bibitem{note-2}
For the result of $N=40$, our numerical accuracy
is not high enough to allow us to precisely
determine the (resonant) transmission peaks,
owing to the extremely dense energy levels
(with actually $500\times 40$ energy levels of the system).
This makes the SOC($\alpha$)-modulation curve of $N=40$
in Fig.\ 4 not very satisfactory. But this imperfect result
may not affect the main underlying physics (effect).
\end{references}
\end{document}